\documentstyle[12pt,aaspp4]{article}

\def\etal{{\it et~al.\ }}

\begin{document}

\title{Surface Brightness of Starbursts at Low and High Redshifts}

\author{Daniel W. Weedman}
\affil{Department of Astronomy and Astrophysics,
525 Davey Laboratory,
The Pennsylvania State University,
University Park, PA 16802}
\authoremail{weedman@astro.psu.edu}

\author{Jeffrey B. Wolovitz}
\affil{Department of Astronomy and Astrophysics,
525 Davey Laboratory,
The Pennsylvania State University,
University Park, PA 16802}
\authoremail{jbw122@psu.edu}

\author{Matthew A. Bershady}
\affil{Department of Astronomy,
475 North Charter St.,
University of Wisconsin,
Madison, WI 53706}
\authoremail{mab@astro.wisc.edu}

\author{Donald P. Schneider}
\affil{Department of Astronomy and Astrophysics,
525 Davey Laboratory,
The Pennsylvania State University,
University Park, PA 16802}
\authoremail{dps@astro.psu.edu}

\begin{abstract}

Observations in the rest frame ultraviolet from various space missions
are used to define the nearby starburst regions having the highest
surface brightness on scales of several hundred pc.  The bright limit
is found to be $6\times10^{-16}$ ergs cm$^{-2}$ s$^{-1}$ \AA$^{-1}$ \
arcsec$^{-2}$ for rest frame wavelength of 1830 \AA. Surface
brightness in the brightest pixel is measured for 18 galaxies in the
Hubble Deep Field having $z > 2.2$. After correcting for cosmological
dimming, we find that the high redshift starbursts have intrinsic
ultraviolet surface brightness that is typically four times brighter
than low redshift starbursts. It is not possible to conclude whether
this difference is caused by decreased dust obscuration in the high
redshift starburst regions or by intrinsically more intense star
formation.  Surface brightness enhancement of starburst regions may be
the primary factor for explaining the observed increase with redshift
of the ultraviolet luminosity arising from star formation.

\end{abstract}


\section{Introduction}

Many new observational results make possible investigations of the 
evolution of star formation within galaxies, extending to redshifts 
approaching four (Madau \etal 1996, Lilly \etal 1996, Connolly \etal 
1997).  To date, the essence of these results is that the global star 
forming rate, deduced from the ultraviolet luminosity integrated over all 
observable galaxies in co-moving volumes of the universe, was 
significantly greater at earlier epochs than it is today. These results 
describe the overall rate of star formation.  This star formation is in 
``starbursts'' for those galaxies whose appearance is dominated by 
short-lived populations of young stars.  While counts of such 
star-forming galaxies can determine global rates, these results do not 
address the question of whether the local characteristics of the star 
formation process within individual galaxies also depends on epoch in the 
universe.  Understanding these characteristics of star formation is 
crucial to explaining the changing rate of star formation with epoch. 

Recent dramatic progress in identifying starburst galaxies at redshifts 
above two in the Hubble Deep Field (Steidel \etal 1996, Lowenthal \etal 
1997) has produced a sample of such galaxies that can be examined with 
sufficient spatial resolution that individual starburst regions can be 
examined within the galaxies. For the first time, it becomes possible to 
compare such regions with those of similar size in nearby galaxies and to 
make this comparison in the rest-frame ultraviolet luminosity, which 
originates completely from the starburst.   This allows an initial 
comparison of the starburst process between galaxies separated in time by 
amounts exceeding two-thirds the age of the universe.

\section{Surface Brightnesses of Starburst Regions}

Our objective in the present analysis is a direct comparison between
the surface brightness of starburst regions at low redshift and at
high redshift, as observed in the rest frame ultraviolet where the
luminosity is completely dominated by the recently formed stars. If
the starburst region is resolved, surface brightness is a very useful
parameter for comparing regions of greatly differing redshift; surface
brightness, unlike luminosity or diameter, does not depend on
cosmological parameters, i.e.  H$_0$, $\Omega_0$ or $\Lambda_0$.  This
makes possible a comparison of local physical conditions that is
independent of cosmological assumptions.

In Euclidean space, surface brightness is independent of distance.
For significant redshifts, surface brightness changes dramatically but
only in terms involving $(1+z)$. When measured in $f_\lambda$ units at
a given rest-frame wavelength, distant objects of the same intrinsic
surface brightness should have their observed surface brightness fade
by $(1+z)^{-5}$ for Friedmann cosmologies.  There are five factors of
$(1+z)$ when observations are in $f_\lambda$ units; two factors
because of the change in unit area observed from the galaxy, one for
change in unit time, one for change in photon energy, and one for
change in unit wavelength bandpass. (If observations are in
bolometric quantities, such as $\lambda f_\lambda$ or $\nu f_\nu$
units, only the first four factors of $(1+z)$ enter. This is equivalent
to observing in a band-pass fixed in either the observed or rest
frame. If counting photons, there is still one less factor of
$(1+z)$.)

Achieving a surface brightness measurement requires that the starburst 
regions be resolved.  We refer to these resolved sources as ``extended 
starburst regions''.  We avoid starbursts associated with galactic nuclei, 
because the starburst luminosity in such cases may be confused with that 
of an unresolved active galactic nucleus. An empirical upper limit to the 
rest frame ultraviolet surface brightness of such regions within nearby 
galaxies is deduced by searching for those nearby extended starburst 
regions with the highest surface brightness.  Observations in the 
ultraviolet from a variety of space missions are the source of the 
relevant data.  These results are summarized in Table 1.  The galaxies 
observed with the various missions were chosen as sources expected to be 
ultraviolet-bright. While it is possible that even brighter extended 
starburst regions have been overlooked, the similarity in the results 
from independent selections of galaxies gives reasonable confidence that 
a meaningful upper limit for surface brightness of extended starburst 
regions in nearby galaxies can be determined from these various 
observations. 

An extensive compilation of observations of star-forming galaxies with
the International Ultraviolet Explorer (IUE) is given by Kinney
et. al.  (1993). For many such observations, the starburst regions are
significantly smaller than the $10^{\prime\prime} \times
20^{\prime\prime}$ IUE aperture, so surface brightnesses within this
aperture are underestimates (ultraviolet images of some nearby
galaxies from the HST Faint Object Camera are in Maoz et al. 1996).
Not surprisingly, the brightest such galaxies within an aperture this
large are generally also the closest, for which the starbursts would
be resolved by the IUE aperture.  Excluding NGC 1068, the five
galaxies with brightest total flux at 1900 \AA \ (their 1863-1963 \AA
\ bandpass) are NGC 1705, 3310, 4449, 5236, and 5253.  The average
surface brightness of these five, with the IUE aperture taken as 200
arcsec$^2$, is $5\times10^{-16}$ ergs cm$^{-2}$ s$^{-1}$ \AA$^{-1}$ \
arcsec$^{-2}$.  Defining the scale size by the $20^{\prime\prime}$
length of the aperture, these brightnesses arise from regions of
average size 700 pc.  While it is arbitrary to restrict our selection
to the five brightest, it can be seen from Table 1 that any other
entries would be more than a factor of two fainter than the brightest
of all (NGC 5236).

The galactic disk with highest observed ultraviolet surface brightness
is the inner disk of NGC 1068, characterized by extensive star
formation throughout the 3 kpc diameter disk centered on the active
nucleus of this prototype Seyfert 2 galaxy.  Numerous individual
bright regions were imaged with the Ultraviolet Imaging Telescope on
the Astro 1 mission (Neff \etal 1994).  The brightest of these is
their knot complex labeled ``region J'', stated to have flux at 1500 \AA
\ of $2\times10^{-14}$ ergs cm$^{-2}$ s$^{-1}$ \AA$^{-1}$.  From their
published image, this region appears dominated by bright knots
covering an area of about 30 arcsec$^2$, which yields a surface
brightness of $7\times10^{-16}$ ergs cm$^{-2}$ s$^{-1}$ \AA$^{-1}$ \
arcsec$^{-2}$, within a characteristic diameter of about 350 pc
($5^{\prime\prime}$ at an assumed distance of 15 Mpc for H$_0 = 75$ km
s$^{-1}$ Mpc$^{-1}$; this value of Ho is assumed throughout).

Several starburst galaxies were imaged at 2200 \AA \ with the Faint
Object Camera of the Hubble Space Telescope (Meurer et. al.  1995 ).
Among these is the highly luminous and well studied starburst galaxy
NGC 3690 (Markarian 171).  The brightest uv knot, NGC 3690-BC, has
surface brightness of $5\times10^{-16}$ ergs cm$^{-2}$ s$^{-1}$
\AA$^{-1}$ \ arcsec$^{-2}$ within diameter of $6.4^{\prime\prime}$
(1360 pc at distance 44 Mpc).

For comparisons to follow with starburst regions at high redshift, we
desire a normalizing wavelength of 1830 \AA.  Nominal corrections to the
observed values in Table 1 are applied by assuming a common spectral
shape for all objects observed.  Taking the median starburst
ultraviolet spectral shape from Kinney \etal (1993) of $f_\lambda
\propto \lambda^{-1}$, the rest-frame surface brightnesses at 1830 \AA
\ are also tabulated.  The results are adequately consistent to define
a meaningful measure of the surface brightness for the brightest local
starbursts. Using these results, we adopt a value of $6\times10^{-16}$
ergs cm$^{-2}$ s$^{-1}$ \AA$^{-1}$ \ arcsec$^{-2}$ for the surface
brightness of the ``brightest'' starburst regions within nearby starbursts
when observed at a rest frame wavelength of 1830 \AA. We will use this
number with which to compare the observed surface brightness of
systems at high redshift.

The image quality of HST allows star forming regions to be resolved,
and surface brightnesses to be measured, even to the highest
observable redshifts.  In the Hubble Deep Field (Williams \etal 1996
), the ``drizzled add'' pixels are $0.04^{\prime\prime}$ in size.
Such a pixel corresponds to a physical size of about 300 pc at $2 < z
< 3$ (for $\Omega_0 = 0.2$; angular sizes change little with redshift
for this $\Omega_0$ and $\Lambda_0 = 0$).  This is sufficiently small
compared to the size of luminous starburst regions in nearby galaxies
(Table 1) that it is meaningful to compare surface brightnesses of
distant starbursts in the HDF with those in nearby galaxies.  Although
spatial resolution in the HDF as defined by the point spread function
is larger than the size of a single pixel, the surface brightness
measures from a single pixel are accurate measures if starburst
regions are fully resolved in the HDF.  If the starburst regions are
unresolved, a measure of the surface brightness in the brightest pixel
provides a lower limit to the actual surface brightness of the source
within an area of 0.016 arcsec$^2$. The surface brightness would be
higher if the source is smaller than a single pixel and is unresolved.
In such cases, however, the starburst regions at high redshift would
be significantly smaller than most of the nearby regions in Table 1.
Because of smearing by the point spread function, measuring surface
brightness in the brightest pixel does not yield a value that is
significantly greater than an average taken over a few pixels. The
brightest pixel is typically found to be a factor of 1.09 times
brighter than the average of the four brightest pixels (600 pc at $2 <
z < 3$) and 1.22 times brighter than the average of the nine brightest
(900 pc at $2 < z < 3$). These apertures are comparable to the IUE,
HST FOC, and UIT apertures used to measure the surface brightness of
the local starbursts listed in Table 1.

There are 18 galaxies in the HDF with known redshifts sufficiently high 
that rest-frame flux measurements are available for ultraviolet 
wavelengths comparable to those at which surface brightnesses of local 
starburst regions have been measured. For these galaxies in the HDF,  the 
ultraviolet surface brightness at 1830 \AA \ rest-frame wavelength is given 
in Table 2. Objects in this Table have redshifts determined by the 
authors referenced, using the Keck Telescope. Images of these objects 
from the HDF are shown in Figure 1. Surface brightness is measured from 
the single brightest pixel using a linear interpolation between the 
effective wavelengths of the two filters which flank the redshifted 1830 
\AA \ wavelength in the observer's frame.  The brightest pixel is chosen from 
the F814W image, and the same pixel is used in the F606W or other images 
to determine the interpolation. The  choice of 1830 \AA \ for the comparison 
wavelength is made so that the object of highest redshift in the analysis 
has rest wavelength corresponding to an observed wavelength no longer 
than the effective wavelength of the F814W image. 

\section{Comparing Low Redshift and High Redshift Starburst Regions}

Ultraviolet observations of nearby starburst regions summarized in \S
2 indicate that the maximum surface brightness of such regions is
about $6\times10^{-16}$ ergs cm$^{-2}$ s$^{-1}$ \AA$^{-1}$ \
arcsec$^{-2}$ when observed at a wavelength of 1830 \AA.  Distant
objects of the same intrinsic surface brightness should have their
observed surface brightness, at the same rest-frame wavelength,
reduced to $6\times10^{-16}$ $(1+z)^{-5}$ ergs cm$^{-2}$ s$^{-1}$
\AA$^{-1}$ \ arcsec$^{-2}$.

Figure 2 displays the observed surface brightnesses at the rest frame
wavelength of 1830 \AA \ (observer's wavelength of $1830(1+z)$ \AA) from the
HDF starburst regions in Table 2.  These are compared to the expected
surface brightness from the scaling from low redshift starbursts.  As
expected from the cosmological fading, surface brightnesses of high
redshift starbursts are dramatically less than for starburst regions
at low redshifts.  Nevertheless, objects in the HDF with high
redshifts have rest-frame surface brightnesses significantly greater
than expected using the scaling from local starbursts.  From Figure 2,
the median excess is a factor of 4, and the brightest starburst in the
Figure exceeds the expected surface brightness {\it by a factor of 11}.
Keeping in mind that these values are only lower limits if starburst
regions are smaller than a single pixel, this result clearly
demonstrates that galaxies at high redshift detected in the HDF have
higher ultraviolet luminosity per unit area than the brightest known
examples of local starbursts.

This is a different conclusion from that of Meurer \etal (1997) and 
Pettini \etal (1998).  These authors conclude from ultraviolet surface 
brightnesses that starburst regions have similar star formation intensity 
per unit area regardless of redshift.  Pettini \etal state that 
starburst regions at high redshift are ``spatially more extended versions 
of the local starburst phenomenon.'' Differences between our conclusions 
and theirs may arise primarily because we measure surface brightnesses within 
the smallest observable spatial scales, using the closest starburst 
regions for comparisons and using only starbursts from the HDF to deduce 
the surface brightness of the starburst regions at high redshift.  Their 
results refer to spatial scales at high redshift averaged over half-light 
radii of about 2000 pc, which are significantly larger scales than those 
within which we measure surface brightnesses.  

The difference in our comparison between low redshift and high
redshift starburst regions is not explainable simply as a selection
effect arising from comparing significantly different volumes of
space.  If a much larger co-moving volume of the universe were
observed to locate starbursts in the HDF compared to the volume of
space used to locate the low redshift comparison starbursts, it might
be expected that rarer, brighter regions would define the brightest
examples in the larger volume.  Consider the volume examined to find
the high redshift starbursts in the HDF.  There are 18 galaxies in
Table 2, distributed roughly evenly in redshift for $2.2 < z < 3.4$,
and 17 of these galaxies exceed the surface brightness of the local
starburst galaxies.  The three WF CCDs of the HDF examined to find
these galaxies cover $1.5 \times 10^{-3}$ deg$^2$.  Within this area
of the sky , the co-moving volume increment in this redshift range is
$15 \times 10^4$ Mpc$^3$ (for $\Omega_0 = 0.2$) or $5 \times 10^4$
Mpc$^3$ (for $\Omega_0 = 1$). These alternative volumes yield volume
densities of the brightest high redshift starbursts between $11 \times
10^{-5}$ Mpc$^{-3}$ and $34 \times 10^{-5}$ Mpc$^{-3}$.

It is more difficult to estimate precisely the low-redshift volume
which has been included to define the low redshift surface brightness,
but that volume can be bracketed reasonably well.  The most distant
galaxy used to define the maximum local starburst surface brightness
is NGC 3690, early identified as the most dramatic example of an
extended starburst in the Markarian sample of galaxies (Gehrz \etal
1983).  Taking the Markarian survey as covering 0.25 of the sky and
taking the distance of NGC 3690 as 42 Mpc, the volume of the universe
examined to find NGC 3690 is $7.5\times10^4$ Mpc$^3$. The next closest
object in Table 1 is NGC 1068, at 15 Mpc distance.  The entire
universe has been examined for bright galaxies within that distance,
so $1.4\times10^4$ Mpc$^3$ has been searched.  It is reasonable to
conclude that these limits ($1.4\times10^4$ Mpc$^3$ to $7.5\times10^4$
Mpc$^3$) bracket the local volumes searched to define the maximum
surface brightness of nearby extended starburst regions as summarized
in Table 1. Based on the 18 sources found in the HDF, we would then
expect $7^{+12}_{-6}$ sources of comparable surface brightness as the
HDF high redshift starbursts in the surveyed local volume. In other
words, local surveys {\it do} cover sufficient volume to find the
types of high surface-brightness starbursts found in the HDF {\it if}
they are at comparable space densities locally.  Using only the 7
objects in Table 1 to define the density of locally-brightest
starbursts, the local density of these starbursts is between
$9\times10^{-5}$ Mpc$^{-3}$ and $50\times10^{-5}$ Mpc$^{-3}$. Because
this density range is very similar to that of the (much more luminous)
high redshift sample, we also can conclude that extended starburst
regions of similar co-moving density have much higher surface
brightness at high redshift compared to low redshift.

The result that starburst regions are significantly brighter per unit 
area at high redshift compared to low redshift is empirical and is our 
primary conclusion.  Understanding why this is the case will require 
knowing more about the spectral characteristics of starburst regions at 
high redshift, especially including infrared luminosities.  There are two 
possibilities for explaining this result.  The first is that starbursts 
at high redshift are intrinsically more intense per unit area than are 
nearby starbursts.  The second possibility is that high redshift 
starbursts are less absorbed by dust, so that a higher proportion of the 
rest frame ultraviolet escapes and the starbursts simply appear brighter 
when observed in the rest frame ultraviolet.  

Meurer \etal (1997) and Pettini \etal (1998) conclude that high
redshift starburst regions are significantly absorbed, although they
differ in conclusions regarding the amount of absorption.  Neither
suggests a systematic difference between low redshift and high
redshift objects, which would discount differing obscuration as the
explanation for the differential brightening which we measure.  Their
efforts to understand the dust absorption in high redshift starbursts
were motivated by the importance of understanding the overall rate of
star formation at high redshift.  This rate can be significantly
underestimated if the ultraviolet light of young stars is heavily
obscured. Both analyses estimate dust obscuration by the reddening
imposed on the ultraviolet spectrum. A difficulty in using only
ultraviolet spectra to determine dust obscuration is that all stars
which contribute to the ultraviolet luminosity must be assumed to be
reddened by an amount small enough that they remain visible but large
enough that the effects of reddening are noticeable in the spectrum.
Stars which are so obscured that negligible ultraviolet luminosity
escapes would not be accounted for.  Because of this effect, estimates
of obscuration derived strictly from ultraviolet spectra provide only
upper limits for the escaping fraction of ultraviolet luminosity.
Accounting for completely obscured stars requires observations in the
rest-frame infrared, where the radiation emerges as re-radiation from
the obscuring dust (see Rowan-Robinson \etal 1997 and Smith \etal
1998).  Such observations are not available for high redshift
starbursts.

The results from Meurer \etal yield a distribution of dust obscuration
in high redshift starburst regions (their Figure 8), a distribution
that can be expressed in terms of the fraction of intrinsic
ultraviolet luminosity which escapes. For the present analysis, we
wish to know only if this distribution is systematically different
compared to low redshift starburst regions.  For the low redshift
starbursts, we can examine the quantitative obscuration by comparing
ultraviolet and infrared luminosities, which together should account
for all of the starburst, regardless of the degree of obscuration.
This analysis is similar to that in Weedman (1991) but with much
improved data. If an IMF is assumed for a starburst and used with
stellar atmospheric models describing the ultraviolet luminosity and
bolometric luminosity of stars of various masses, the ratio of
intrinsic ultraviolet to bolometric luminosity is known for the
starburst.  For the present analysis, we use the model described in
Meurer et al (1997), which normalizes such that $f_{\lambda 2200} =
1.5\times10^{-4} f_{\rm bol}$.

Because of the availability of far-infrared fluxes from IRAS and
ultraviolet fluxes from IUE, a substantial number of starburst
galaxies have measured bolometric fluxes (Schmitt \etal 1997).  As
these authors comment, there are uncertainities in comparing $f_{uv}$ and
$f_{\rm bol}$ because of differing aperture sizes in the observations, but the
IUE aperture is large enough that it generally includes all of the
relevant starburst , and the IRAS fluxes, though having poorer spatial
resolution, are generally dominated by the same starburst regions
(Calzetti \etal 1995).  In Table 3 are given all of the starburst
galaxies in Schmitt \etal (1997), including those which they classify as both
low reddening and high reddening, listing the $f_{\rm bol}$ as read from their
plots.  Also given are the $f_{\lambda 2200}$ fluxes as deduced from
the summary of IUE observations in Kinney et al (1993).  (The
$f_{\lambda 2200}$ are deduced by interpolating between the entries
with central wavelengths of 1913 \AA \ and 2373 \AA.)  The intrinsic
$f_{\lambda 2200}$ deduced from $f_{\rm bol}$ using the Meurer result that
$f_{\lambda 2200} = 1.5\times10^{-4} f_{\rm bol}$ is also listed in Table
3.  Comparing observed and intrinsic $f_{\lambda 2200}$ yields the
fraction of ultraviolet luminosity which escapes, given in the Table.

In Figure 3, the cumulative distributions of escaping fraction for the
ultraviolet luminosity are shown.  For the median in the distribution
of low redshift starbursts from Table 3, only 10\% of the intrinsic
2200 \AA \ luminosity escapes.  This is similar to the results deduced
by Meurer \etal (1997) using only the shape of the ultraviolet continuum as
an indicator of dust absorption.  For their high redshift sample of
galaxies, the cumulative distribution is also shown in Figure
3. Comparison of the distributions gives no indication that a
systematically larger fraction of ultraviolet luminosity emerges
unobscured from the high redshift starburst regions.

Recall that the obscuration deduced for the high redshift sample is
based only on the ultraviolet spectra and could not account for
completely obscured stars.  The distribution for high redshift,
therefore, actually shows upper limits for the escaping fraction of
ultraviolet luminosity; the real fraction could be smaller if there
are undetected stars because of obscuration.  Even so, these upper
limits are comparable to the values for the low redshift sample, which
{\it do} account for completely obscured stars.  Any systematic
difference between the samples arising from the different methods used
to estimate obscuration would have underestimated obscuration for the
high redshift starburst regions.  This is counter to the possible
explanation being considered, which requires that high redshift
starburst regions have less, not more, obscuration compared to low
redshift regions.

The conclusions above indicate that differing dust obscuration does not 
account for the systematically different surface brightnesses between low 
redshift and high redshift starburst regions.  This is not, however, 
fully consistent with results reported by Pettini \etal (1998). Although 
they do not present an analysis of individual spectra, Pettini \etal 
conclude that a higher fraction of ultraviolet luminosity escapes in the 
high redshift galaxies than was concluded by Meurer \etal (1997). They 
report a median dust correction to the ultraviolet luminosity of a factor 
of 3, although they state it could range from 2 to 6.  This dust 
correction is substantially less than the factor of approximately 10 
shown by the distributions in Figure 3. If it is correct that as much as 
one-third of intrinsic ultraviolet luminosity characteristically escapes 
from high redshift starburst regions, this would enhance their surface 
brightness almost enough to account for our observed difference between 
surface brightnesses of low redshift and high redshift regions (factor of 
4 in the median).  We note, however, that the Pettini \etal results also can only 
describe an upper limit for escaping ultraviolet luminosity (equivalent 
to a lower limit to the obscuration factor) because of the absence of any 
information on bolometric luminosities.  

\section{Summary}

By examining available ultraviolet observations of low redshift
starburst regions, we define an upper limit to the surface brightness
of such starburst regions as observed on scales of several hundred pc.
This limit is $6\times10^{-16}$ ergs cm$^{-2}$ s$^{-1}$ \AA$^{-1}$ \
arcsec$^{-2}$ for rest frame wavelength of 1830 \AA. Ultraviolet
surface brightnesses at comparable scales for high redshift starburst
regions ($z > 2.2$) are measured from galaxies in the Hubble Deep
Field.  After correcting for cosmological effects, we find that high
redshift starbursts have intrinsic ultraviolet surface brightness that
is typically four times brighter than low redshift starbursts.

Because of differing conclusions about the amount of obscuration in high 
redshift starburst regions,  we are left with an ambiguous interpretation 
as to whether the unexpectedly high surface brightness of starburst 
regions at high redshift is caused by diminished dust obscuration.  If 
not, some effect is required to make high redshift star formation 
intrinsically more intense per unit area than star formation in the 
nearby universe.  Either way, our results indicate that something was 
different on a local scale within starburst regions at redshifts above 
two compared with those in nearby galaxies, making star formation appear 
brighter per unit area of a starburst region.   

This increased surface brightness may be a significant factor in 
explaining the increasing luminosity density with redshift observed for 
ultraviolet-bright galaxies.  The distribution with redshift of 
star-forming galaxies observed in ground-based surveys and in the HDF 
indicates that the star formation rate per co-moving volume peaks at 
redshift about unity and remains 5 to 10 times greater than in the nearby 
universe for $z > 2$ (Madau \etal 1996, Lilly \etal 1996, Connolly \etal 
1997).  If there is no change in the size or co-moving density of 
starburst regions, but the star formation intensity per region increases 
a factor of four, this would be a major factor in explaining the observed 
differences between the nearby universe and the high redshift universe.  
In this scenario, the early universe did not require significantly more 
starburst regions than are observed in the nearby universe,  but each 
region had much more intense (or much less obscured) star formation.  Our 
result is not adequate to prove such a scenario, because this would 
require that the surface brightness enhancement apply to all starburst 
regions, whereas our result is able to show this enhancement for only the 
brightest regions in both nearby and high redshift samples. 

\acknowledgments DPS acknowledges support from NSF grant
AST-9509919. MAB acknowledges support from NASA grant NAG5-6032.

\clearpage
\begin{figure}
\plotfiddle{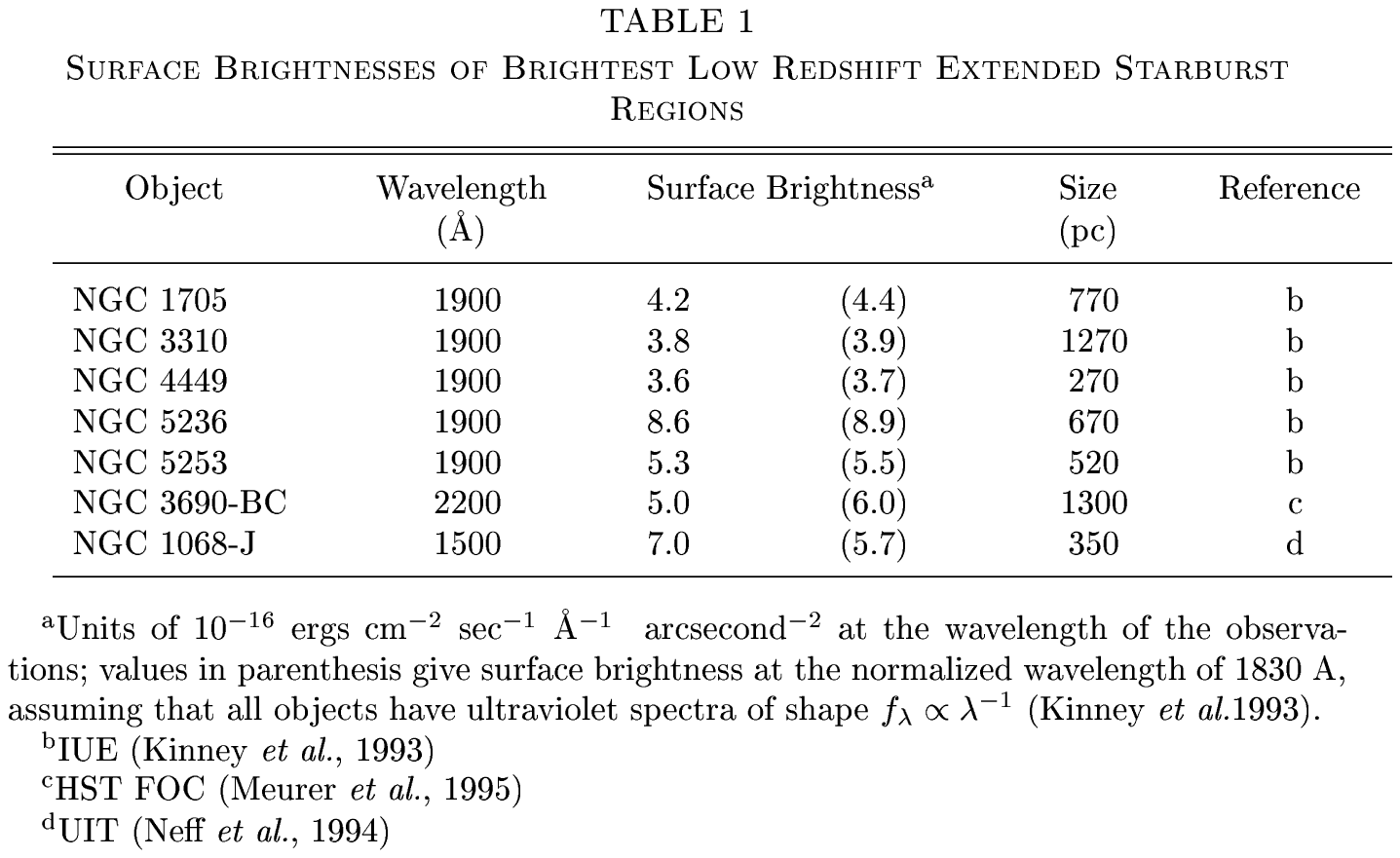}{8in}{0}{100}{100}{-300}{-100}
\end{figure}

\clearpage
\begin{figure}
\plotfiddle{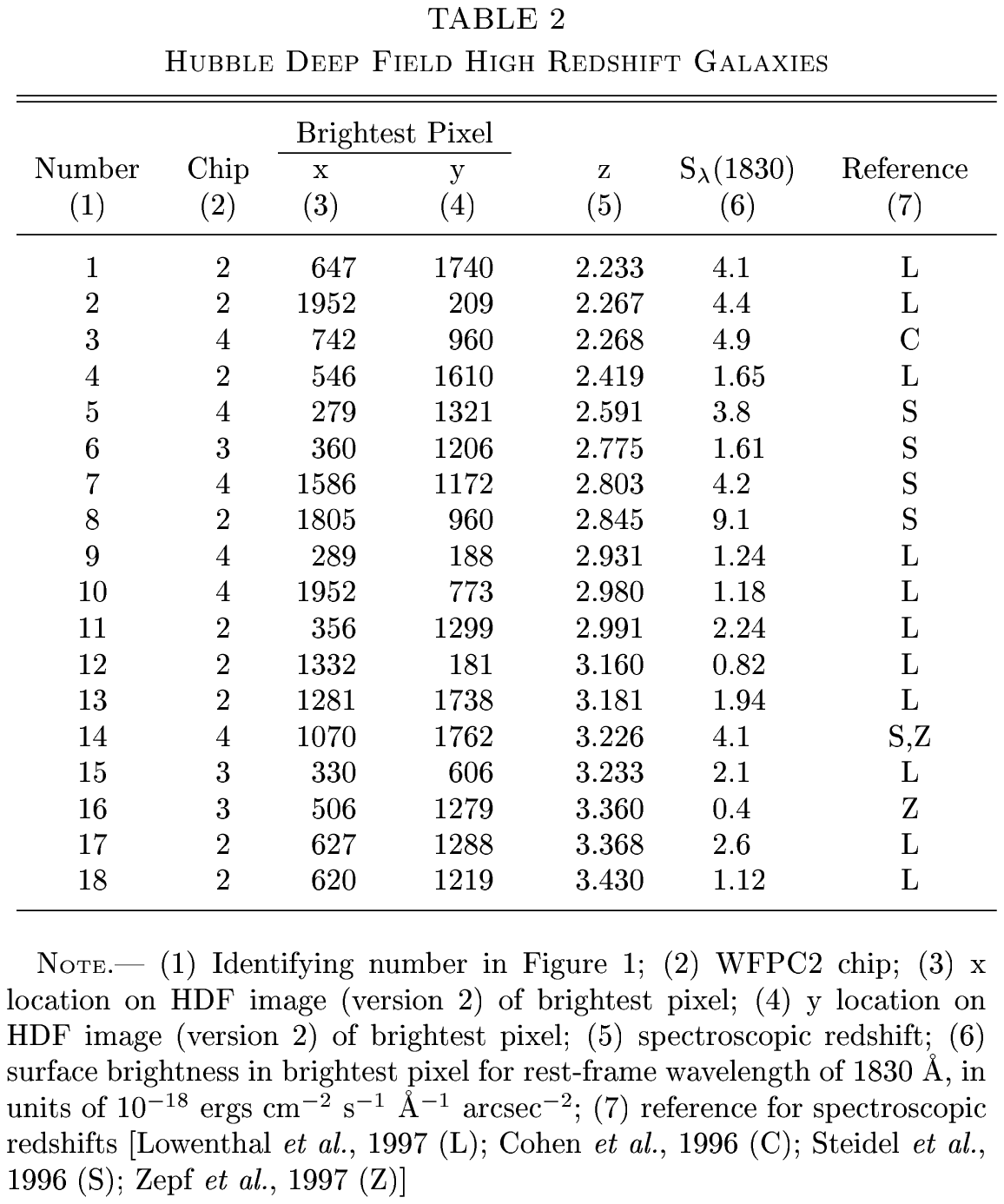}{8in}{0}{100}{100}{-300}{-100}
\end{figure}

\clearpage
\begin{figure}
\plotfiddle{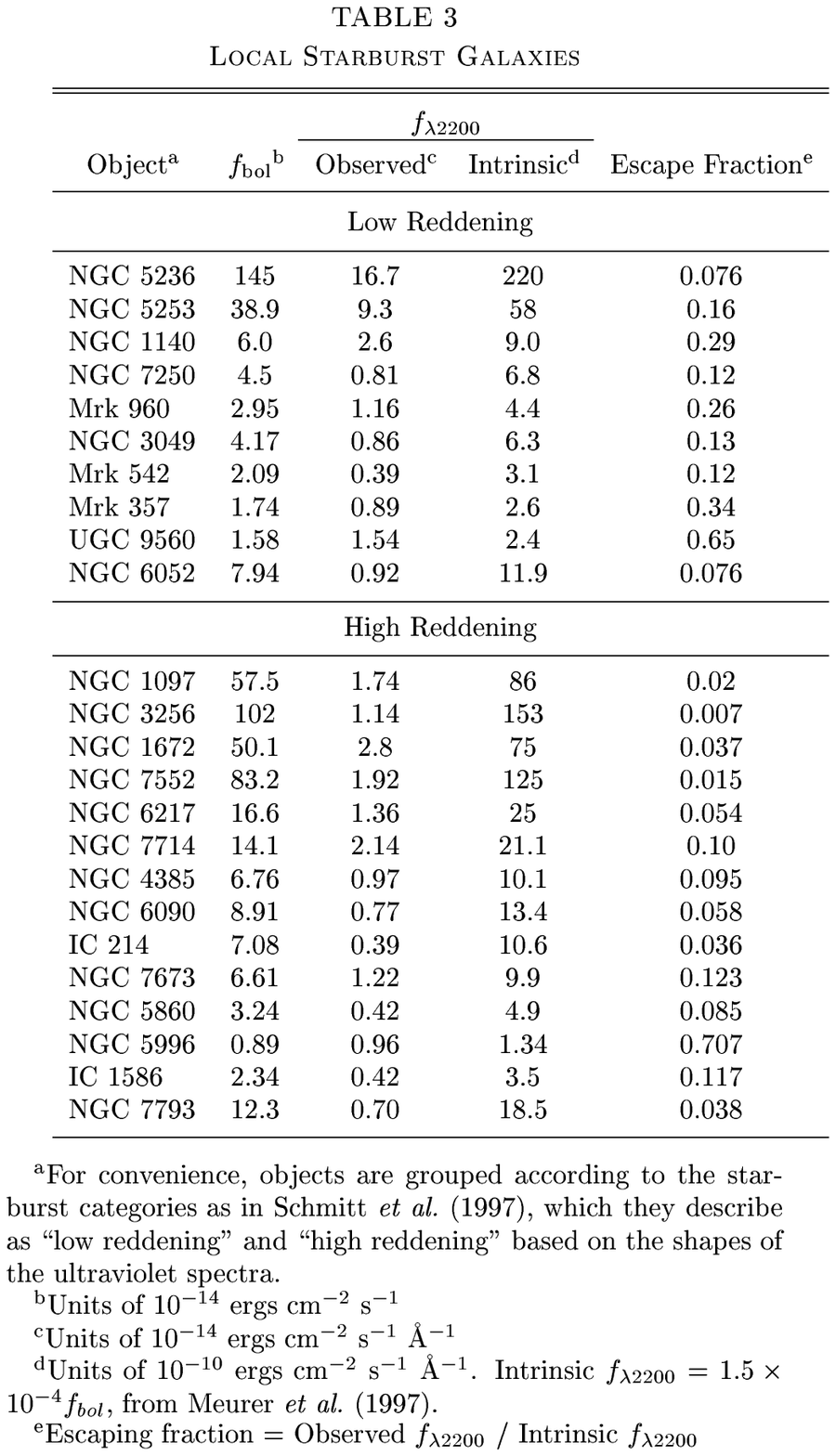}{8in}{0}{100}{100}{-300}{-100}
\end{figure}

\clearpage

\title {Figure Captions}

\figcaption[]{Images of the galaxies in Table 1, from the F814W images
of the Hubble Deep Field. Each box is $3.32^{\prime\prime} \times
3.32^{\prime\prime}$; each tick mark corresponds to roughly 900 pc
(H$_0 = 75$ km s$^{-1}$ Mpc$^{-1}$ and $\Omega_0 =
0.2$). \label{fig1}}

\figcaption[]{Surface brightness (units of $10^{-18}$ ergs cm$^{-2}$
s$^{-1}$ \AA$^{-1}$ \ arcsec$^{-2}$) at 1830 \AA \ compared to
$(1+z)$. Filled triangles: observed values for galaxies in Table 2.
Solid curve: expected values for brightest local starburst regions
($6\times10^{-16}$ ergs cm$^{-2}$ s$^{-1}$ \AA$^{-1}$ \ arcsec$^{-2}$,
see text), faded by cosmological factor of $(1+z)^{-5}$.\label{fig2}}

\figcaption[]{Normalized cumulative distribution of ultraviolet
luminosity which escapes starburst region without being absorbed by
dust. Solid line: low redshift starburst regions from Table 3 (24
objects total). Dashed line: high redshift starburst regions from
Meurer \etal (1997, 23 objects total). \label{fig3}}

\clearpage

\begin{figure}
\plotfiddle{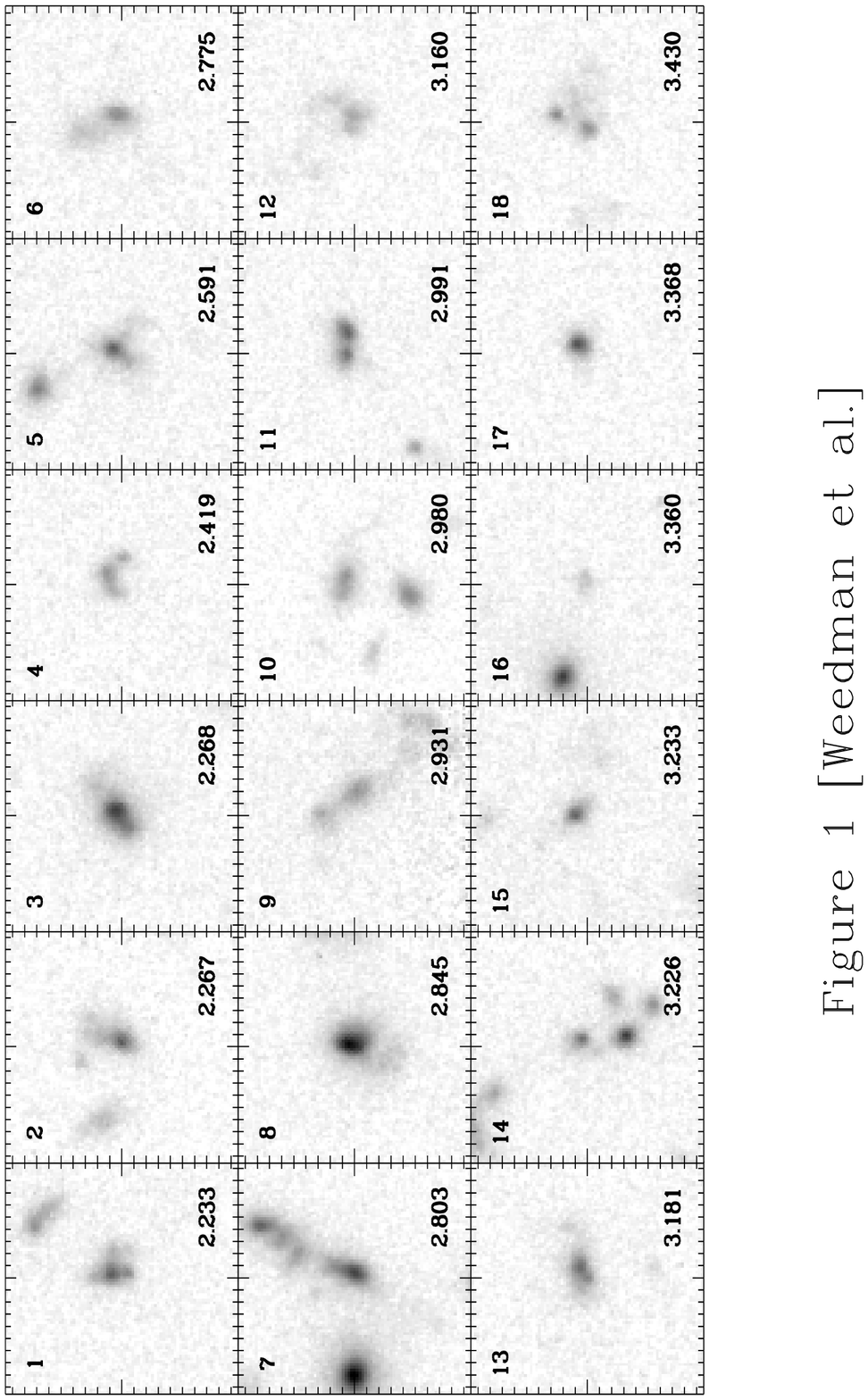}{8in}{0}{100}{100}{-300}{-115}
\end{figure}

\begin{figure}
\plotfiddle{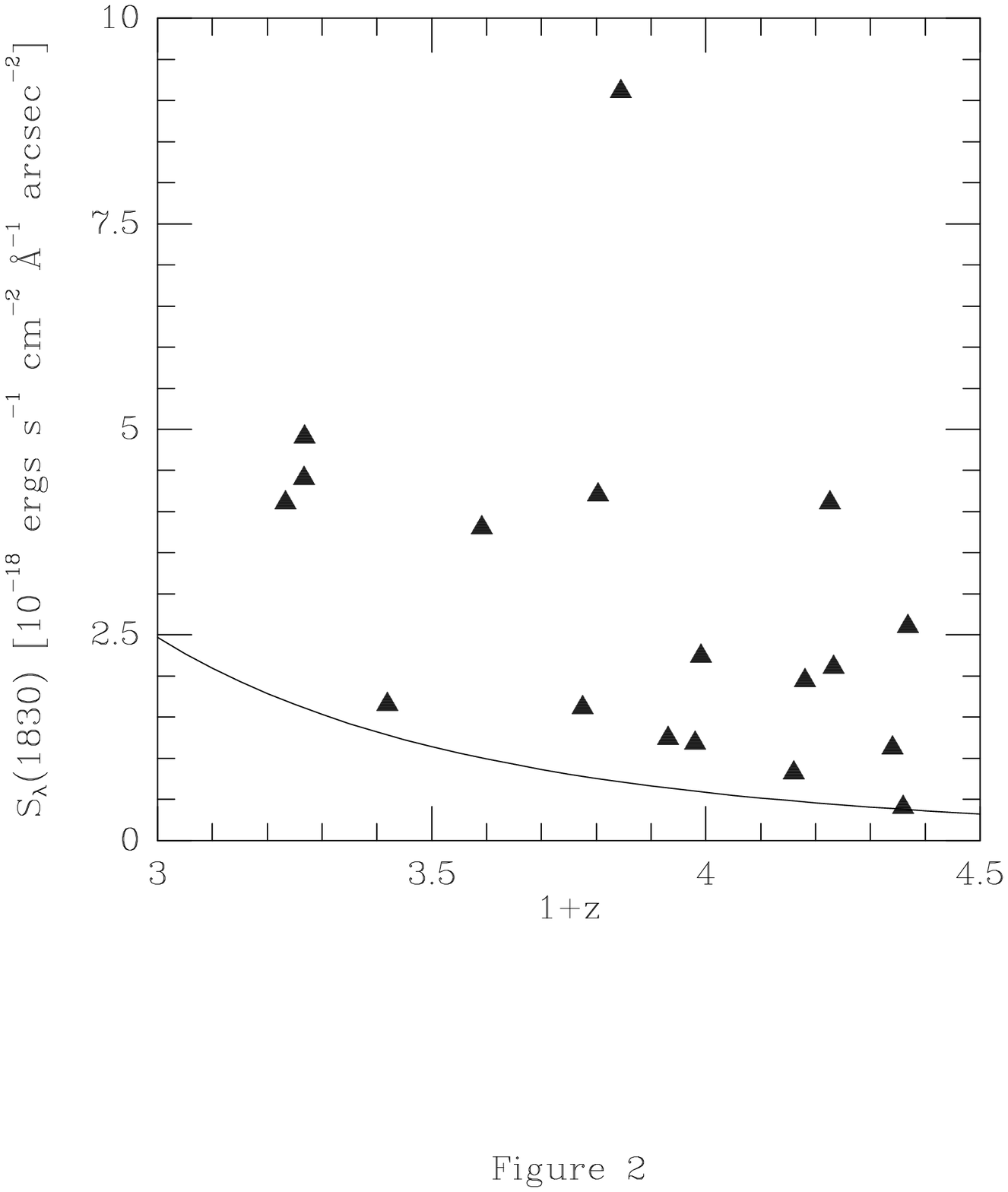}{8in}{0}{100}{100}{-300}{-100}
\end{figure}

\begin{figure}
\plotfiddle{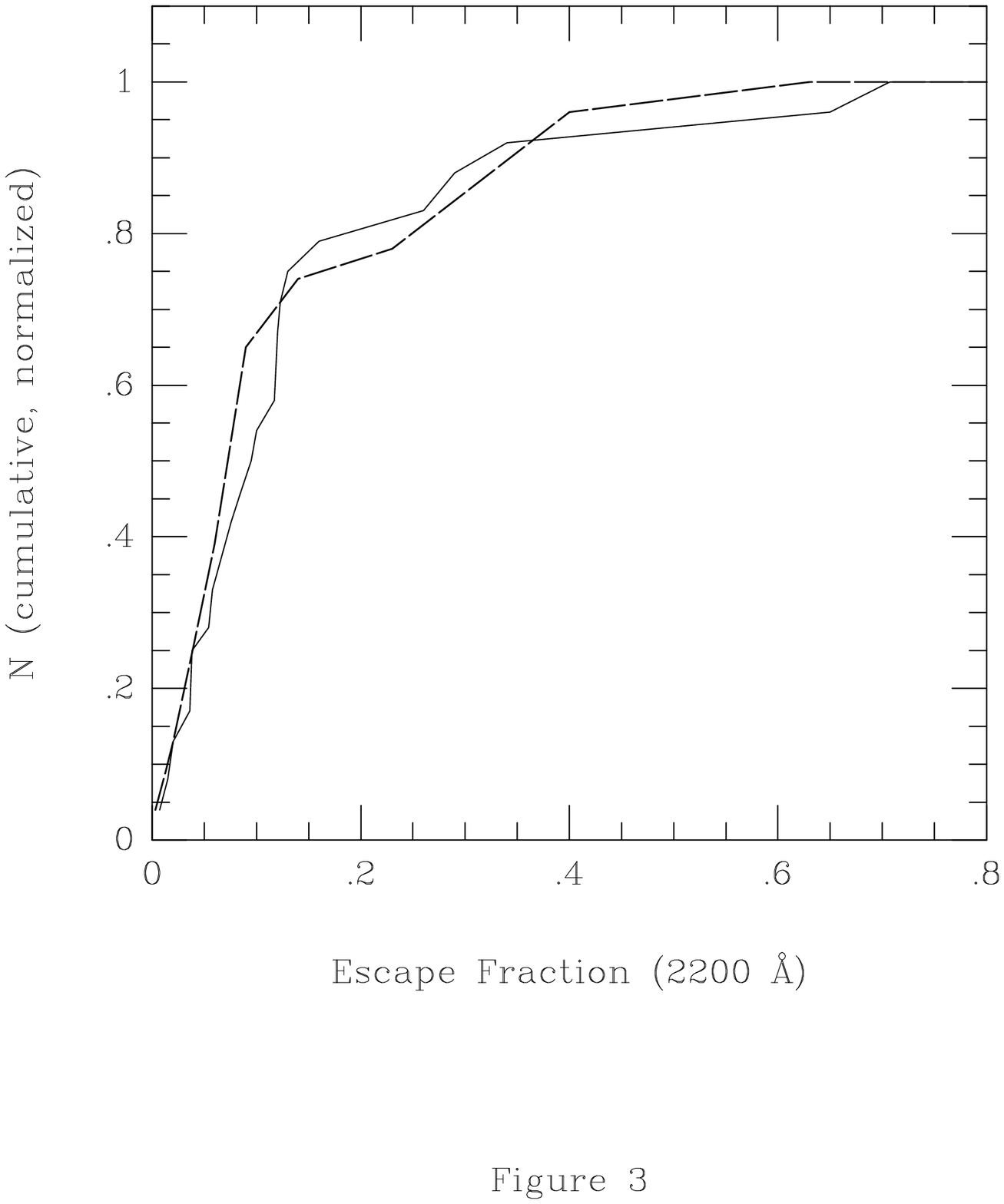}{8in}{0}{100}{100}{-300}{-100}
\end{figure}

\end{document}